\newcommand{\half}{\frac{1}{2}}

 \documentstyle[preprint,aps,tighten]{revtex}
\begin{document}
\draft
\title{Transverse Radiation realized as Deformed Harmonic Oscillators}
\author{ P. Narayana Swamy }
\address{Department of Physics, Southern Illinois University,
Edwardsville IL 62026}

\maketitle
\begin{abstract}
We present a theory of quantized radiation fields described in
terms of q-deformed harmonic oscillators. The creation and
annihilation operators satisfy deformed commutation relations and
the Fock space of states is constructed in this formalism in terms
of basic numbers familiar from the theory of quantum groups.
Expressions for the Hamiltonian and momentum arising from deformed
Heisenberg relations are obtained and their consequences
investigated. The energy momentum properties of the vacuum state
are studied.  The commutation relation for the fields is shown to
involve polarization sums more intricate than those encountered in
standard quantum electrodynamics, thus requiring explicit
representations of polarization vectors. The electric field
commutation rules are investigated under simplifying assumptions
of polarization states, and the commutator in the deformed theory
in this case is shown to be reminiscent of the coordinate-momentum
uncertainty relation in the theory of q-deformed quantum
oscillators.
\end{abstract}

 \vspace{2.8in}
Electronic address:   pswamy@siue.edu\\

\indent August 2004


\section{Introduction}
The task of quantizing transverse electromagnetic radiation is
facilitated by first establishing that the system of pure
radiation fields is equivalent to a set of harmonic oscillators.
As is well-known, this is accomplished by demonstrating that the
Hamiltonian of the radiation fields is  of the same mathematical
form as the Hamiltonian of a set of harmonic oscillators
\cite{Heitler}. Other quantities such as the momentum of radiation
fields can also be cast in this form. The quantum theory of
radiation is then straightforward and follows the steps of
quantizing the system of harmonic oscillators. In other words the
fundamental premise of the quantum theory of radiation has to do
with the equivalence of transverse radiation and a set of
non-interacting harmonic oscillators.

On the other hand we have gained a great deal of insight into the
theory of q-deformed harmonic oscillators from the wisdom of
decades of investigation \cite{Biedenharn,Macfarlane,PNS} of a
theory of the system, built on a Fock space of states arising from
modified commutation relations for the annihilation and creation
operators. Deformed oscillators are thus realized as quantum
groups. The physical meaning of such a deformation can be
interpreted as a modification of the Planck constant
\cite{Biedenharn}. In other words, the q-deformation can be
regarded as a modification of the Planck constant just as quantum
mechanics might be regarded as a deformation of the classical
system.

We may accordingly pose the following question. If the standard
quantum theory of transverse electromagnetic radiation is
completely equivalent to a system of harmonic oscillators, what
kind of transverse radiation fields would correspond to a system
of q-deformed harmonic oscillators? The purpose of this work is to
formulate such a theory of pure transverse electromagnetic
radiation and display its salient properties. This work is
certainly not yet a complete investigation of all aspects of the
theory but rather a modest, preliminary study to highlight some of
the prominent features. This paper is organized as follows. Sec.
II is devoted to the formulation of the theory of electromagnetic
radiation that would correspond to a set of q-deformed harmonic
oscillators. This formalism employs basic numbers.  Sec.III  deals
with the Hamiltonian of the system expressed in terms of
annihilation and creation operators obeying modified commutation
relations. We also  describe the construction of Fock space of
photon states. In Sec.IV we investigate the momentum of the
radiation fields. We also study the properties of the vacuum
state. Sec.V contains a useful representation of the polarization
vectors needed for our investigation. The rationale for this is as
follows. This formalism is more complicated than standard quantum
electrodynamics and most investigations would require an explicit
representation of the polarization vectors satisfying the usual
properties. In Sec. VI, we study the commutation relations
satisfied by radiation fields, specifically the commutation
relations of the electric field components. We illustrate the main
feature of deformation of field commutation relations by
considering a simplistic example of specific polarization states.
Sec.VII contains a summary and concluding remarks.

\section{The Formalism}

We seek a plane wave representation of potentials $\textbf{A}$,
the conjugate momentum $\textbf{P}$ and fields $\textbf{E},
\textbf{B}$ representing pure transverse radiation which will be
consistent with q-deformed oscillators. We begin with the fact
that the Hamiltonian of the standard quantized radiation fields is
described by
\begin{equation}\label{1}
    H = \int \;d^3r \;  \left \{\, 2 \pi c^2 |\textbf{P}|^2 + \frac{1}{8 \pi}\,
     (\nabla \times \textbf{A})^2 \, \right \}\, ,
\end{equation}
where the momentum conjugate to the potential is
\begin{equation}\label{2}
    \textbf{P}= \frac{1}{4 \pi} \left (\frac{1}{c}
    \frac{\partial {\textbf{A}}}{\partial t}+
    \nabla\phi \right )\, .
\end{equation}
The well-known standard analysis
\cite{Heitler,Landau,Lurie,Trigg,Schiff,BjorkenDrell,Gasiorowicz}
leads to the Hamiltonian of the form
\begin{equation}\label{3}
    H = \sum'_{{\bf k},\lambda} \left( 4\pi c^2\,
    p^{\dag}_{{\bf k},\lambda}\,  p_{{\bf k},\lambda}
    + \frac{k^2}{4 \pi} \, x^{\dag}_{{\bf k},\lambda}\, x_{{\bf k},\lambda}    \right )\, ,
\end{equation}
which is obtained from the expansions
\begin{equation}\label{4}
    \textbf{A}(\textbf{r},t)=\sum_{{\bf k},\lambda}'
    \{x_{{\bf k},\lambda}(t)
     {\bf u}_{{\bf k},\lambda}(\textbf{r})
       + x^{\dag}_{{\bf k},\lambda}(t) {\bf u}^{\ast}_{{\bf k},\lambda}
     (\textbf{r})  \}
\end{equation}
and
\begin{equation}\label{5}
    \textbf{P}(\textbf{r},t)=\sum_{{\bf k},\lambda}'\{p_{{\bf k},\lambda}(t)
     {\bf u}_{{\bf k},\lambda}(\textbf{r})    + p^{\dag}_{{\bf k},\lambda}(t)
     {\bf u}^{\ast}_{{\bf k},\lambda}
     (\textbf{r})  \}\, .
\end{equation}
Here the sum over $\lambda =1,2$ corresponds  to the two
transverse polarization degrees of freedom. The foundations of the
quantum theory of radiation formulated a long time ago
\cite{Heitler} need to be generalized.  We shall adhere to the
notation employed in \cite{Schiff}, which includes many of the
nuances of the analysis. The prime accompanying the expansions
above indicates that the sum is over only half the $\bf k$-space
i.e., positive $\bf k$ direction: this avoids duplication of ${\bf
u}_{-{\bf k},\lambda}$ in ${\bf u}^{\ast}_{{\bf k},\lambda}$. At
the end of the calculation \cite{Schiff}, we shall be able to
replace it by the ordinary sum over all of $\bf k$-space. The
dynamical variables $x_{{\bf k},\lambda}$ and $p_{{\bf
k},\lambda}$, the coordinates and momenta,  are formal objects
introduced so that the form of the Hamiltonian, Eq.(\ref{3}),  is
that of harmonic oscillators. The plane wave representation is
necessary to express the potentials and fields in a complete
orthonormal set of plane waves according to
\begin{equation}\label{6}
    {\bf u}_{{\bf k},\lambda}(\textbf{r})= L^{-3/2}\;
    \mbox{\boldmath $\varepsilon$}_{{\bf k},\lambda}\; e^{i \bf{k}\cdot
    \bf{r}}\, .
\end{equation}
Thus for each momentum $\textbf{k}$ and polarization
$\lambda=1,2,$ \,  the plane wave is constructed and employed in
the expansions in Eqs.(\ref{3}-\ref{5}) and $\mbox{\boldmath
$\varepsilon$}_{{\bf k},\lambda} $ represents the polarization
vector describing transverse radiation for a given momentum and
polarization direction.

The objects $x_{{\bf k},\lambda} \, {\bf u}_{{\bf k},\lambda}$ and
$p_{{\bf k},\lambda}\, {\bf u}_{{\bf k},\lambda} $ in
Eqs.(\ref{4},\ref{5}) can be expressed as linear combinations of
plane waves $e^{i({\bf{k}}\cdot {\bf{r}} \, \mp \, kct)}$ with
coefficient operators $a_{{\bf k},\lambda}$ and $a'^{\dag}_{{\bf
k},\lambda}$ in the undeformed theory. In the deformed theory,
however, these annihilation and creation operators play subtler
roles, as we shall see.

Accordingly, we shall now proceed to introduce the basic premises
of the formulation that will lead to the realization of transverse
radiation as q-deformed harmonic oscillators.

\indent 1. First, the canonical commutation relations satisfied by
the  dynamical variables $x_{{\bf k},\lambda}$ and $p_{{\bf
k},\lambda}$ in standard electrodynamics are
\begin{equation}\label{7}
    [x_{{\bf k},\lambda},p^{\dag}_{{\bf k}',\lambda'} ]\; = \;
    [x^{\dag}_{{\bf k},\lambda},p_{{\bf k}',\lambda'}] \; = \;
    i\,  \hbar \delta_{{\bf k} {\bf k}'}\delta_{\lambda \lambda'}
\end{equation}
which must now be modified by the following commutation relations
valid for the q-deformed system:
\begin{equation}\label{8}
[x_{{\bf k},\lambda},p^{\dag}_{{\bf k}',\lambda'} ]\; = \;
    [ x^{\dag}_{{\bf k},\lambda},p_{{\bf k}',\lambda'} ] \; = \;
    i\,  \hbar \delta_{{\bf k} {\bf k}'}\delta_{\lambda \lambda'}\;
    ([N_{{\bf k},\lambda} + 1]- [N_{{\bf k},\lambda}])\, ,
\end{equation}
where $[a,b]= a b - b a $. The commutator itself thus remains
unmodified and the right hand side displays the effect of
deformation. The bracket numbers on the right hand side here
denote the basic numbers defined by
\begin{equation}\label{9}
    [x]= \frac{q^x - q^{-x}}{q-q^{-1}},
\end{equation}
where $q$ is the parameter (base)  signifying deformation. The
bracket number or the basic number $[N]$ is not the number
operator but $N$ is. In this formulation \cite{Exton} which is
symmetric under $q \longrightarrow q^{-1}$, it suffices to
restrict its range to $0 < q < 1$ with no loss of generality.
There are many useful properties of bracket numbers known in the
literature, such as the following representation:
\begin{equation}\label{}
    [x] = \frac{\sinh x \gamma}{\sinh \gamma}, \quad  q=
    e^{\gamma}\, .
\end{equation}

\indent 2. The time dependence of operators is governed by
Heisenberg equations of motion,
\begin{equation}\label{11}
    i\hbar \frac{\partial}{\partial t} \,  x_{{\bf k},\lambda}(t) =
    [x_{{\bf k},\lambda}(t), H]= 4 \pi i \hbar
    c^2 p_{{\bf k},\lambda}\; ,
\end{equation}
and
\begin{equation}\label{12}
i\hbar \frac{\partial}{\partial t} \, p_{{\bf k},\lambda}(t) =
[p_{{\bf k},\lambda}(t), H]=
 -\frac{i\hbar k^2}{4 \pi} x_{{\bf k},\lambda} \; ,
\end{equation}
where $k = |{\bf k}|$. There is thus no deformation introduced in
the time dependence. We could have introduced deformation of the
equations of motion governing time dependence here by employing
the Jackson derivative, playing an important role leading to
q-calculus \cite{ALPNS1}, which is known to preserve the whole
structure of thermostatistics. However, we shall refrain from such
a generalization here.

\indent 3. For the deformed oscillators, $[N_{{\bf k},\lambda}]$
is not the number operator; instead the bracket numbers and the
annihilation and creation operators are related by
\begin{equation}\label{10}
    [N_{{\bf k},\lambda}]= \frac{k}{2 \pi \hbar c}
    a^{\dag}_{{\bf k},\lambda} a_{{\bf k},\lambda}\; ; \quad
[N_{{\bf k},\lambda}+1]= \frac{k}{2 \pi \hbar c}
    a_{{\bf k},\lambda} a^{\dag}_{{\bf k},\lambda}\, .
\end{equation}
Our analysis below will enable us  to unambiguously  establish the
above identification.

We are now ready to derive the consequences of the  basic premises
introduced above. Eqs.(\ref{11},\ref{12}) immediately lead to
\begin{equation}\label{14}
    \frac{\partial^2}{\partial t^2} \,  x_{k,\lambda}(t) = -k^2 c^2 x_{k,\lambda}(t)\,,
\end{equation}
and a similar differential equation for $p_{k,\lambda}(t)$.
Consequently, we obtain the solutions in the form of the linear
combination
\begin{equation}\label{15}
    x_{{\bf k},\lambda}(t)= a_{{\bf k},\lambda}\, e^{-ikct} +
a'^{\dag}_{{\bf k},\lambda}\, e^{ikct}
\end{equation}
 and
\begin{equation}\label{16}
p_{{\bf k},\lambda}(t)= -\frac{ik}{4 \pi c} \left \{a_{{\bf
k},\lambda}\, e^{ikct} - a'^{\dag}_{{\bf k},\lambda}\,
e^{ikct}\right \}\,,
\end{equation}
just as in the undeformed theory \cite{Schiff}. At the end of the
calculation, we shall be able to identify $a'_{{\bf k},\lambda}$
with $a_{-{\bf k},\lambda}$. Inverting Eqs. (\ref{15},\ref{16}),
we obtain the relations
\begin{eqnarray}
 \nonumber a_{{\bf k},\lambda} &=& \half \{ x_{{\bf k},\lambda}+
 \frac{4 \pi ic} {k}
 p_{{\bf k},\lambda}\}\,  e^{i k c t}\, ,  \\
  a'^{\dag}_{{\bf k},\lambda} &=&
   \half \{ x_{{\bf k},\lambda}- \frac{4 \pi ic} {k}
 p_{{\bf k},\lambda}\} \, e^{-i k c t}\,\label{A1} \, ,
\end{eqnarray}
and taking the hermitian conjugate leads to the further relations
\begin{eqnarray}
 \nonumber a^{\dag}_{{\bf k},\lambda} &=&
 \half \{ x^{\dag}_{{\bf k},\lambda}
 -  \frac{4 \pi ic} {k}
 p^{\dag}_{{\bf k},\lambda}\} \, e^{-i k c t} \, , \\
  a'_{{\bf k},\lambda} &=&
  \half \{ x^{\dag}_{{\bf k},\lambda} + \frac{4 \pi ic} {k}
 p^{\dag}_{{\bf k},\lambda}\} \, e^{i k c t}\label{A2}\, .
\end{eqnarray}
 We can now evaluate the commutator, and after some algebra,  obtain
\begin{equation}\label{21}
    [a_{{\bf k},\lambda}, \; a^{\dag}_{{\bf k}',\lambda'}]=
    \frac{2 \pi \hbar c}{k}\delta_{{\bf k}
     {\bf k}'}\, \delta_{\lambda \lambda'}
    \left ([N_{{\bf k},\lambda} +1]- [N_{{\bf k},\lambda}]   \right )\, ,
\end{equation}
upon utilizing Eq.(\ref{8}). In the limit $q \rightarrow 1$ this
goes over to the standard commutation relation since
$[N_{k,\lambda} +1]- [N_{k,\lambda}] \rightarrow 1$ for all ${\bf
k},\lambda $. Similarly, we obtain from the second pair of
equations,
\begin{equation}\label{22}
[a'_{{\bf k},\lambda}, \; a'^{\dag}_{{\bf k}',\lambda'}]=
    \frac{2 \pi \hbar c}{k}\delta_{{\bf k} {\bf k}'}\, \delta_{\lambda \lambda'}
    \left ([N'_{{\bf k},\lambda} +1]- [N'_{{\bf k},\lambda}]   \right )\, .
\end{equation}
We observe that the above equations evidently confirm the
relations contained in Eq.(\ref{10}).

\section{The Hamiltonian}

The Hamiltonian given by Eq.(\ref{3}), together with (\ref{8})
describes q-deformed oscillators. We can evaluate it in terms of
the creation and annihilation operators given by
Eqs.(\ref{A1},\ref{A2}) and express the result  as
\begin{equation}\label{23}
    H= \sum'_{{\bf k},\lambda}\frac{k^2}{2 \pi}
     \left(a_{{\bf k},\lambda} a^{\dag}_{{\bf k},\lambda}
     +     a^{\dag}_{{\bf k},\lambda} a_{{\bf k},\lambda}  \right )\, .
\end{equation}
where the summation is over half the ${\bf k}$-space (positive
$\bf {\bf k} $ direction). Utilizing Eq.(\ref{10}), the
Hamiltonian can be cast in the form
\begin{equation}\label{24}
 H= \sum'_{{\bf k},\lambda} \hbar c k \left (
 [N_{{\bf k},\lambda} + 1] + [N'_{{\bf k},\lambda}]   \right  )\, .
\end{equation}
Identifying $N'_{{\bf k},\lambda}$ in terms of negative  $\bf k$
direction,  we may now extend the sum over all of ${\bf k}$-space,
thus removing the prime in the sum \cite{Schiff},  and obtain the
result
\begin{equation}\label{25}
H= \sum_{{\bf k},\lambda}\half \hbar c k \left (
 [N_{{\bf k},\lambda} + 1] + [N_{{\bf k},\lambda}]   \right  )\, ,
\end{equation}
where the bracket numbers appear instead of the number operator
for $q\neq 1$. This reduces to the standard result $\sum \hbar c k
(N_{{\bf k},\lambda} + \half )$ in the limit $q \rightarrow 1$ of
of the undeformed theory. Consequently, the result in
Eq.(\ref{25}) includes the zero point energy.

 Finally, we may collect all the commutation
relations in the form
\begin{eqnarray}
  \nonumber [x_{{\bf k},\lambda}, \, p^{\dag}_{{\bf k},\lambda}] &=&
  [x^{\dag}_{{\bf k},\lambda}, \, p_{{\bf k},\lambda}]=
   i \hbar \left ( [N_{{\bf k},\lambda} + 1] - [N_{{\bf k},\lambda}]\right ) \\
  \label{26}[x^{\dag}_{{\bf k},\lambda},\,  p_{{\bf k},\lambda}] &=& [x_{{\bf k},\lambda},
  \,  p^{\dag}_{{\bf k},\lambda}]=
   i \hbar
  \left ( [N'_{{\bf k},\lambda} + 1] - [N_{{\bf k},\lambda}]\right
  )\, ,
\end{eqnarray}
where, as before, the prime on $N_{{\bf k},\lambda}$ can be
identified with negative directions of $\bf k$.

It is important to observe that the deformed Heisenberg
commutation relations for every ${\bf k}$ and $\lambda$ as in
Eq.(\ref{8}) directly lead to the identification in Eq.(\ref{10}).
Our analysis further shows that this leads to the Hamiltonian,
Eq.(\ref{25}), expressed in terms of the basic numbers, which are
again essential in describing the deformed oscillators. We thus
conclude that the Hamiltonian of transverse electromagnetic
radiation, realized as deformed oscillators, is a direct
consequence of  the following ingredients of the theory: the
q-deformed Heisenberg commutation relations for the dynamical
variables $x_{{\bf k},\lambda} \, ,\; p_{{\bf k},\lambda} $; the
expressions for $a_{{\bf k},\lambda}, a^{\dag}_{{\bf k},\lambda}$
in terms of the dynamical variables $x_{{\bf k},\lambda}, p_{{\bf
k},\lambda}$; the time dependence given by the Heisenberg
equations of motion.

We may arrive at the result, Eq.(\ref{25})  directly from the
expression for the energy of radiation fields, Eq.(\ref{1}), as an
instructive exercise; or equivalently from
\begin{equation}\label{}
    H= \frac{1}{8 \pi}\,\int \;d^3r \;  \left ( |E|^2 + |B|^2 \right ) \,.
\end{equation}
It suffices to consider the contribution from the electric field
alone since that would be half the total. We then determine after
a straightforward  calculation:
\begin{eqnarray}
  \nonumber H &=& \frac{1}{8 \pi}\, \int \, d^3r\,
   E^i({\bf r},t) \,   E^i({\bf r},t) \\
   \nonumber &=& \frac{1}{8 \pi}\, \int \, d^3r
   \, L^{-3}\sum'_{{\bf k},\lambda}
   \sum'_{{\bf k}',\lambda'}\, (k \, k') \, \varepsilon^i_{{\bf k},\lambda}
    \, \varepsilon^i_{{\bf k}',\lambda'} \\
  \nonumber & & \times \left \{   a_{{\bf k},\lambda}
  \, e^{i({\bf k}.{\bf r} - k c t)}\,   - a^{\dag}_{{\bf k},\lambda}
   \, e^{-i({\bf k}.{\bf r} - k c t)}\,  \right \} \\
  \nonumber   & & \times \left \{a^{\dag}_{{\bf k}',\lambda'}
   \, e^{-i({\bf k}'.{\bf r} - k' c t)}\,
   - a_{{\bf k}',\lambda'}\,
    e^{i({\bf k}'.{\bf r}- k' c t)}\,
  \right \} \\
   &=& \frac{1}{8 \pi} \sum_{{\bf k}, \lambda}\, k^2
   (a_{{\bf k},\lambda}\,  a^{\dag}_{{\bf k},\lambda}
    + a^{\dag}_{{\bf k},\lambda}\,  a_{{\bf k},\lambda})\, .
\end{eqnarray}
Hence we obtain the desired result
\begin{equation}\label{}
 H = \, \frac{1}{8 \pi}\,\int \;d^3r \;  |E|^2
  \,= \,\sum_{{\bf k},\lambda} \, \frac{1}{4} \hbar
c k (\, [N_{{\bf k}, \lambda} +1] + [N_{{\bf k}, \lambda}]\, )\,,
\end{equation}
where the sum is now over all of $\bf k$ space.

 We shall now complete this
section by discussing the Fock space of the deformed harmonic
oscillators. We begin by observing that while the commutation
relation such as in Eq.(\ref{21}) is most convenient and natural,
 we can easily derive an  alternative form. From Eq.(\ref{10}),
we immediately derive
\begin{equation}\label{}
     a_{{\bf k}, \lambda} a^{\dag}_{{\bf k}, \lambda} -
    q \, a^{\dag}_{{\bf k}, \lambda}a_{{\bf k}, \lambda} = q^{-N}\, ,
\end{equation}
which is  the preferred form standard in the literature. We
introduce the Fock states  $|n_{{\bf k}, \lambda } \rangle$ as
follows.  By analyzing the action on this state by operators
$a_{{\bf k}, \lambda}$ and $a^{\dag}_{{\bf k}, \lambda}$ we find
\begin{equation}\label{}
    a^{\dag}_{{\bf k}, \lambda} |n_{{\bf k}, \lambda }
    \rangle = \sqrt{[n_{{\bf k},\lambda}+1]}\;
    |n_{{\bf k}, \lambda } \rangle; \;\;
a_{{\bf k}, \lambda} |n_{{\bf k}, \lambda } \rangle =
\sqrt{[n_{{\bf k},\lambda}]} \;
    |n_{{\bf k}, \lambda }-1 \rangle \, .
\end{equation}
Consequently we obtain, for any ${\bf k}, \lambda$,
\begin{equation}\label{}
|n_{{\bf k}, \lambda }  \rangle = \frac{(a^{\dag}_{{\bf k},
\lambda } )^{n_{{\bf k}, \lambda }}} {\sqrt{[n_{{\bf k}, \lambda}]
!}}\; |0 \rangle \, ,
\end{equation}
where
\begin{equation}\label{}
    [r]! = [r]\, [r-1]\, \cdots 1 \, .
\end{equation} Accordingly we may build the general Fock state by
\begin{equation}\label{}
   |\, n_{{\bf k}_1, \lambda_1 },\; n_{{\bf k}_2,
   \lambda_2 } \cdots \,
    \rangle = \left ( \prod_{s}
    \frac{ (a^{\dag}_s)^{n_s
    }}
    {\sqrt {[n_s]!} } \right ) \; |0 \rangle  \, ,
\end{equation}
where the abbreviation $s$ denotes ${\bf k}_s, \lambda_s$.

\section{Momentum of Radiation}

Starting from the definition of the momentum of the radiation
fields
\begin{equation}\label{27}
    \textbf{G} = \frac{1}{4 \pi  }\int \;d^3 r \; \left (\textbf{E} \times
    \textbf{B}  \right )\, ,
\end{equation}
we  may express it in terms of the dynamical variables as
\begin{equation}\label{28}
\textbf{G} = i \sum'_{{\bf k},\lambda} \textbf{k}\, (p_{{\bf
k},\lambda} \, x^{\dag}_{{\bf k},\lambda} - p^{\dag}_{{\bf
k},\lambda}\, x_{{\bf k},\lambda})\, .
\end{equation}
We shall now utilize Eqs.(\ref{15},\ref{16}) in order to express
the momentum in terms of the creation and annihilation operators.
Noting that $[a_{{\bf k},\lambda},\, a_{{\bf k},\lambda}]= [
a^{\dag}_{{\bf k},\lambda}, \, a^{\dag}_{{\bf k},\lambda} ]=0$ and
making use of the commutation relations, Eq.(\ref{21}, \ref{22}),
we obtain
\begin{equation}\label{29}
    \textbf{G}=  \frac{1}{4 \pi c}\sum'_{{\bf k},\lambda}
     k\,  \textbf{k}
    \{( a_{{\bf k},\lambda} \, a^{\dag}_{{\bf k},\lambda} + a^{\dag}_{{\bf k},\lambda} \,
    a_{{\bf k},\lambda} )- (a'_{{\bf k},\lambda}\,  a'^{\dag}_{{\bf k},\lambda} +
     a'^{\dag}_{{\bf k},\lambda}\,  a'_{{\bf k},\lambda})  \}
\end{equation}
which can be readily expressed in the form
\begin{equation}\label{30}
\textbf{G}=  \half \sum'_{{\bf k},\lambda} \hbar \textbf{k}
    \{[N_{{\bf k},\lambda}+1]+ [N_{{\bf k},\lambda}]-  ([N'_{{\bf k},\lambda}+1]+ [N'_{{\bf k},\lambda}])       \}\, .
\end{equation}
First, it is instructive study the limit $q\rightarrow 1$ in the
undeformed theory. In this case the bracket numbers become
ordinary numbers and the terms containing unity cancel. We can
then express it in terms of the sum over the full ${\bf k}$-space
in this case and  obtain the familiar, standard result,
\begin{equation}\label{31}
\textbf{G}=   \sum_{{\bf k},\lambda} \hbar \textbf{k}
   \,  N_{{\bf k},\lambda}       \, .
\end{equation}
It is important to note that the zero point contribution cancels
out and the sum above is over all of ${\bf k}$. Now returning to
the case of $q\neq 1$, and identifying $N'_{{\bf k},\lambda}$ as
belonging to negative ${\bf k}$, proceeding as before, we obtain
the result
\begin{equation}\label{32}
\textbf{G}=  \half \sum_{{\bf k},\lambda} \hbar \textbf{k}
   \left ([ N_{{\bf k},\lambda}+1]+ [N_{{\bf k},\lambda}]\right )       \}\,.
\end{equation}
Hence we conclude that when $q\neq 1$, the zero point contribution
does not trivially vanish.  On the other hand, by comparing with
the Hamiltonian, we find that the operator energy-momentum
relation $H=c|G|$ is satisfied. Consequently, the expression for
momentum of radiation is given by the two different expressions,
Eqs.(\ref{31}, \ref{32}) for $q=1$ and $q\neq 1$ and it would be
misleading to take the limit directly in Eq.(\ref{32}). It is
interesting to note that in the undeformed case of $q=1$, the
limit of Eq.(\ref{32}) appears to be in agreement with the result
in \cite{Landau}. However it must be stressed that the above
results are correct only in their respective domain.

In view of the discussion above, it may be worthwhile to inspect
the results more closely in order to remove any ambiguity.
Eq.(\ref{30}), may be alternatively re-expressed as
\begin{equation}\label{33}
    \textbf{G}= \half \sum'_{{\bf k},\lambda}\, \hbar {\bf k}
     \left \{
    q[N_{{\bf k},\lambda}] + q^{-N_{{\bf k},\lambda}}
    +[N_{{\bf k},\lambda}] - (q[N'_{{\bf k},\lambda}]
     + q^{-N'_{{\bf k},\lambda}} + [N'_{{\bf k},\lambda}])
     \right
    \}\, .
\end{equation}
since $[N+1]= q [N] + q^{-N}$.  The numbers $N'$ are identified
with negative momentum  values. Finally we can combine into a sum
over the full ${\bf k}$-space and thus obtain
\begin{equation}\label{34}
\textbf{G}= \half \sum_{{\bf k},\lambda}\, \hbar {\bf k} \left(
(q+1)[N]+q^{-N} \right),
\end{equation}
where the sum is without the prime, thus confirming the previous
result in Eq.(\ref{32}). Accordingly we may summarize our results
as follows:

\begin{equation}\label{35}
    \textbf{G}= \left \{ \begin{array}{cc}
    \sum_{{\bf k},\lambda} \, \hbar {\bf k} N_{{\bf k},\lambda}\, ,
     & \mbox{$q=1$}\, , \\
    \\ \sum_{{\bf k},\lambda} \,\half \hbar {\bf k}
    ([N_{{\bf k},\lambda}+1] + [N_{{\bf k},\lambda}])\, ,  & \mbox{$q \neq 1$}
    \end{array}
    \right.
\end{equation}
and
\begin{equation}\label{36}
H= \left \{ \begin{array}{cc}
    \sum_{{\bf k},\lambda} \, \hbar c k (N_{{\bf k},\lambda}+ \half)\,
     , & \mbox{$q=1$} \\
    \\ \half \sum_{{\bf k},\lambda} \, \hbar c k
    ([N_{{\bf k},\lambda}+1]+ [N_{{\bf k},\lambda}])\, , &
    \mbox{$q \neq 1$}
    \end{array}
    \right.
\end{equation}

We conclude this section by discussing the energy and momentum
eigenvalues for the vacuum state, which seems to be of
considerable importance, thus providing an explicit clarification
and reiteration of the above discussion. First we observe that
\begin{equation}\label{A11}
    N\, |\, 0 \rangle =0, \quad [N] \, |\, \,  0 \rangle =0 \,.
\end{equation}
The latter result follows from the  representation $ [N] = \sinh
N\gamma / \sinh \gamma$. In a similar manner, we observe that
\begin{equation}\label{A10}
    (N+1)\, |\,  0 \rangle = | \,0 \rangle ,
    \quad [N+1]\, |\,  0 \rangle = | \,0 \rangle \, .
\end{equation}
As a result,  we are now able to directly examine
Eqs.(\ref{35},\ref{36}) and conclude that for $q=1$, the momentum
of the vacuum state vanishes whereas it has zero point energy; on
the other hand, in the undeformed case, $q \neq 1$, both the
momentum and energy have zero point values.

\section{Polarization Vectors}

In order to proceed with further investigations, it is necessary
to study the polarization vectors describing transverse
electromagnetic radiation in the deformed theory, although this is
a straightforward matter in ordinary electrodynamics. We know that
the polarization vectors $\varepsilon^i_{\lambda}$   are defined
by orthogonal unit vectors
\cite{Heitler,Landau,Lurie,Trigg,Schiff,BjorkenDrell,Gasiorowicz}
satisfying the property of orthonormality
\begin{equation}\label{37}
    \sum_{i=1}^3\varepsilon^i_{\lambda}
     \varepsilon^i_{\lambda'}= \delta_{\lambda
    \lambda'}\, ,
\end{equation}
where  the space and polarization indices  are $i=1,2,3$ and
$\lambda = 1,2$;  and the completeness relation
\begin{equation}\label{38}
    \sum_{\lambda=1,2}\varepsilon^i_{\lambda}
    \varepsilon^j_{\lambda}= \delta_{ij}-
    \frac{k_i k_j}{k^2}\, .
\end{equation}
The completeness property above is rather obvious since the unit
vectors would ordinarily satisfy
\begin{equation}\label{39}
\sum_{\lambda=1}^3\varepsilon^i_{\lambda} \varepsilon^j_{\lambda}=
\delta_{ij}\, ,
\end{equation}
if the polarization were to be summed over $\lambda = 1,2,3$;  we
merely need to introduce $\varepsilon^i_3 = k_i/ |k| $ and
subtract the product $k_i k_j / |k|^2$ from the above to obtain
the sum over $\lambda =1,2$, leading to Eq.(\ref{38}).

 In order to perform many calculations involving the sum
 over polarizations in the deformed theory, such as the evaluation
 of $\sum_{\lambda} \,
 \varepsilon^i_{\lambda} \, {\cal O}_{\lambda}\,
 \varepsilon^j_{\lambda}$,
 we need an explicit representation of the
polarization vectors satisfying the above properties and for this
purpose we proceed as follows.

Given an arbitrary vector $\bf A$,  the  transverse polarization
vectors are easily constructed by the prescription
\begin{equation}\label{40}
   \mbox {\boldmath $\varepsilon $}_1 = \alpha ({\bf A}- \frac{{\bf k}
   (k \cdot A)}{k^2})
    \, , \quad  \mbox {\boldmath $\varepsilon $}_2 = \frac{{\bf k}\times
    \mbox {\boldmath $\varepsilon $}_1}{|k|}=
   \alpha \frac{  {\bf k}  \times {\bf A}  }  {|k|}\, ,
\end{equation}
so that, together with ${\hat{\bf k}}$, they form a right-handed
triad of unit vectors.  We determine the normalization to be
\begin{equation}\label{41}
    \alpha^2= \frac{k^2}  {k^2 A^2 - (k \cdot A)^2}\, ,
\end{equation}
which still provides a generous class of vectors $\bf A$ for our
purpose. It is easily verified that they satisfy the
orthonormality property, Eq.(\ref{37}). Proceeding to confirm the
completeness property we obtain
\begin{equation}\label{42}
\varepsilon^i_{\lambda}
    \varepsilon^j_{\lambda}= \alpha^2 \left (
    A_i - \frac{k_i (k \cdot A)}{k^2}\right )
    \left (
    A_j - \frac{k_j (k \cdot A)}{k^2}\right )+
    \frac{\alpha^2}{k^2}({\bf k}\times {\bf A})_i
    {\bf k}\times {\bf A})_j\, .
\end{equation}
This is difficult to analyze for arbitrary $\bf A$ and hence it is
prudent to choose a convenient vector.

 Choice 1: if we choose
${\bf A}= \hat{\bf x}$, we can immediately verify the
orthonormality property, Eq.(\ref{37}). We also find
\begin{equation}\label{43}
\varepsilon^i_1
    \varepsilon^j_1 + \varepsilon^i_2
    \varepsilon^j_2=
    \alpha^2 (\hat{\bf x}- \frac{{\bf k}k_1}{k^2})_i
(\hat{\bf x}- \frac{{\bf k}k_1}{k^2})_j + \frac{\alpha^2}{k^2}
(k_3\hat{\bf y}- k_2 \hat{\bf z} )_i (k_3\hat{\bf y}- k_2 \hat{\bf
z} )_j\, .
\end{equation}
Explicitly, we can verify the following results
\begin{eqnarray}
  \nonumber \varepsilon^1_1
    \varepsilon^1_1 + \varepsilon^1_2
    \varepsilon^1_2&=& 1-\frac{k_1^2}{k^2}, \quad
    \varepsilon^2_1
    \varepsilon^2_1 + \varepsilon^2_2
    \varepsilon^2_2= 1-\frac{k_2^2}{k^2}, \,
      \\
  \label{44} \varepsilon^1_1
    \varepsilon^2_1 + \varepsilon^1_2
    \varepsilon^2_2 &=&   -\frac{k_1 k_2}{k^2},
     \quad \varepsilon^3_1
    \varepsilon^3_1 + \varepsilon^3_2
    \varepsilon^3_2 = 1-\frac{k_3^2}{k^2}, \; {\, \rm etc}. \, ,
\end{eqnarray}
and thus verify the completeness relation, Eq.(\ref{38}). We shall
have occasion to utilize the above result in the next section.

There could be other choices. We find ${\bf A}=\hat{\bf k} $ is
not desirable since it does not produce a non-zero polarization
vector.

Choice 2: we may take  ${\bf A}=  (\hat{\bf n}\times \hat{\bf k})$
as our second choice, where $\hat{\bf n}$ is an arbitrary unit
vector. We thus have, in this case,
\begin{equation}\label{45}
\mbox {\boldmath $\varepsilon $}_1= \beta \, (\, {\hat{\bf n}
\times \hat{\bf k}}\, ), \quad \mbox {\boldmath $\varepsilon $}_2
= \beta \,
 (\, \hat{\bf n} - \hat{\bf k}\, (\hat{\bf n}\cdot \hat{\bf k}) \, )\, .
\end{equation}
The normalization constant must now be
\begin{equation}\label{46}
    \beta^2=\frac{k^2}{k^2 - (n \cdot k)^2}\, .
\end{equation}This satisfies the orthonormality property.
If we specialize to the case $\hat{\bf n} = \hat{\bf x}$, we
further obtain the result
\begin{equation}\label{47}
    \varepsilon^i_1
    \varepsilon^j_1 + \varepsilon^i_2
    \varepsilon^j_2=
    \frac{\beta^2} {k^2} ( k_2\hat{\bf z}- k_3 \hat{\bf y} )_i
( k_2\hat{\bf z}- k_3 \hat{\bf y} )_j +\beta^2 (\hat{\bf x}-
\frac{{\bf k}k_1}{k^2})_i (\hat{\bf x}- \frac{{\bf
k}k_1}{k^2})_j\, .
\end{equation}
Calculating with the various components, we can again verify the
results stated in Eq.(\ref{44}) and thus verify the completeness
property. We shall also utilize this result in the next section.

\section{Canonical commutation relations for fields}

We shall now investigate the field commutation relations
corresponding to the q-deformed oscillators. In order to highlight
the prominent features, it suffices to consider the commutation
relation of a typical pair such as that of the electric field
components. Employing the expansions
\begin{equation}\label{48}
    E_i({\bf r},t)= L^{-3/2}\sum_{{\bf k}, \lambda}
    (i k) \, \varepsilon^i_{{\bf k},\lambda}\left \{a_{{\bf k},\lambda}
    e^{i(k \cdot r -kct)}-a^{\dag}_{{\bf k},\lambda}
    e^{-i(k \cdot r-kct)}
    \right \}\,
\end{equation}
and
\begin{equation}\label{49}
    E_j({\bf r}',t')= L^{-3/2}\sum_{{\bf k'}, \lambda'}
    (i k') \varepsilon^j_{{\bf k'},\lambda'}
    \left \{a_{{\bf k'},\lambda'}
    e^{i(k' \cdot r'-k'c t')}-a^{\dag}_{{\bf k'},\lambda'}
    e^{-i(k' \cdot r'-k' c t')}
    \right \}\,,
\end{equation}
we derive the result
\begin{equation}\label{50}
    [E_i({\bf r},t), E_j({\bf r}',t')]=
    4 \pi i \hbar c  \, L^{-3}\, \sum_{{\bf k},\lambda}\,  k
    \, \varepsilon^i_{\lambda}\,  \varepsilon^j_{\lambda}\,
    \sin ({\bf k} \cdot \mbox {\boldmath $\rho $}- k c \tau)
    \left     ( \, [N_{{\bf k},\lambda}+1]
     - [N_{{\bf k},\lambda}] \,  \right )\,    .
\end{equation}
where $\mbox{\boldmath $\rho$} = {\bf r}- {\bf r}',\tau = t - t'
$.  It is readily verified that, for the undeformed
case\cite{Schiff}, this reduces  to  :
\begin{equation}\label{51}
    \lim_{q \rightarrow 1}\,  [E_i({\bf r},t), E_j({\bf r}',t')]
    = L^{-3} ( 4 \pi\, i \,\hbar \,c )\, \sum_{{\bf k}, \lambda}k \, \varepsilon^i_{\lambda}
    \varepsilon^j_{\lambda} \, \sin ({\bf k} \cdot
     \mbox {\boldmath $\rho $} - k c \tau)\,
    \,  .
\end{equation}
 To determine this when
$q \neq 1$, we need to evaluate the polarization sum
\begin{equation}\label{52}
    \sum_{\lambda}\varepsilon^i_{\lambda}\,  \varepsilon^j_{\lambda}
\left
    ( [N_{{\bf k},\lambda}+1] - [N_{{\bf k},\lambda}]  \right )\,.
\end{equation}
From Eq.(\ref{42}), this turns out to be
\begin{equation}\label{53}
    \sum_{\lambda}\alpha^2 \,
    \left \{(A_i - \frac{k_i (k.A)}{k^2})(A_j - \frac{k_j (k.A)}{k^2})
    + \frac{1}{k^2} ({\bf k} \times {\bf A})_i
    ({\bf k} \times {\bf A})_j
    \right \} ([N_{{\bf k},\lambda}+1] -[N_{{\bf k},\lambda}] )\,
    .
\end{equation}
This is intractable due to the arbitrariness of $\bf A$ and the
dependence of $[N]$ on $\bf k$ and $\lambda$. We shall consider
the special choice of last section, the first choice, ${\bf A} =
\hat{\bf x}$. Again, in order to study the nature of the
deformation, we shall  also consider only the components
corresponding to $i=1, j=1$. In this case the polarization sum
reduces to
\begin{equation}\label{54}
    \left (\, 1- \frac{k_1^2}{k^2}\, \right) \, (\, [N_{{\bf k},1}+1]-
    [N_{{\bf k},1}]\, )
\end{equation}
since only $\lambda =1$ survives. Replacing the sum by an
integral, we thus determine the commutation relation to be
\begin{equation}\label{55}
[E_1({\bf r},t), E_1({\bf r}',t')]= \frac{1}{(2 \pi)^3}\, \int \;
d^3 k \, \frac{1}{k}\,  ([N_{{\bf k},1}+1]-
    [N_{{\bf k},1}])\,\sin ({\bf k} \cdot \mbox {\boldmath $\rho $}
     - k c \tau) \, .
\end{equation}
 Even after the simplifying  choices  made, the
 modified commutation relation is considerably different from the
 undeformed case.  The deformation is the simplest when
  the number operator is  independent of $\bf k$. In this case, we
  have
  \begin{equation}\label{56}
    ([N_1+1]- [N_1])= \frac{\cosh (N_1 +\half)\gamma}{\cosh \half
    \gamma}\, ,
\end{equation}
and this factor may be taken outside the integral.  If we make the
second choice as in Eq.(\ref{45}), we obtain for the same pair of
components,
\begin{equation}\label{57}
    [E_1({\bf r},t), E_1({\bf r}',t')]=
    \frac{1}{(2 \pi)^3}\, \int
\;  d^3 k \,\frac{1}{k}\,  ([N_{{\bf k},2}+1]-
    [N_{{\bf k},2}])\,\sin ({\bf k} \cdot \mbox
    {\boldmath $\rho $} - k c \tau) \,,
\end{equation}
since the $\lambda =2$ term survives. In such cases, and only
under the special choices, the commutator is scaled by the factor
above which is the same factor that appears in the fundamental
commutation relation \cite{Biedenharn}, the uncertainty relation
of q-deformed oscillators :
\begin{equation}\label{58}
    [x,p]= i \hbar \; (\, [N +1] - [N]\, )= \frac{\cosh (\,N+\half
    \,)\gamma}{\cosh \gamma}\, .
\end{equation}
More generally, the right hand side of Eqs.(\ref{55},\ref{57})
would be rather intricate due to the dependence on $\bf k$ and
$\lambda$ and the derivative operations on the Green function and
subsequent integration would be a formidable task. However, in the
simple circumstance described above, the q-deformation is revealed
by the factor $([N_1+1]- [N_1])$ or $([N_2+1]- [N_2])$.

\section{Concluding remarks}

We have investigated the formalism of quantized radiation fields
 corresponding to a set of q-deformed harmonic oscillators. This
formalism is based on expressing the Hamiltonian of transverse
radiation fields in terms of the dynamical variables, analogous to
coordinates and momenta, which satisfy deformed Heisenberg
uncertainty relations. The resulting quantum theory of  transverse
radiation is accordingly equivalent to q-deformed harmonic
oscillators familiar from quantum group investigations. This
formalism is technically complicated  due to the vector nature of
the radiation fields, with  momentum vector and polarization
vector appearing as degrees of freedom of the oscillators. Our
goal has been to study the salient features of the theory such as
the Hamiltonian, the momentum and canonical commutation relations
for the electric field components without being daunted by the
complications, rather than investigate all the nuances of the
theory.

We have avoided using Jackson derivatives leading to a more
general form of time dependence for the dynamical variables and
concomitantly for the annihilation and creation operators.
Although the details are not explicitly presented in this work, we
have verified that the JD would lead to time dependence via
q-exponential functions $E_q(\pm i k c t)$ instead of the ordinary
exponential functions and the field commutation relations will be
further complicated by dependence on time through q-sine
functions. We have taken the approach requiring the simplest
deformation.

The task of quantizing radiation in the manifestly covariant
formulation is further complicated by the use of Lorentz gauge,
requiring a four dimensional analysis as well as the full use of
gauge invariance. It also requires the use of four component
polarization vectors \cite{Lurie,Gasiorowicz}. This is, however,
unnecessary in the present investigation and we have succeeded in
employing a simpler approach by working in the radiation gauge so
that we restrict ourselves to transverse polarizations.

 The complication of the vector
nature of the fields is such that even with two polarization
directions, many calculations would involve polarization sums
including projection operators. This problem is not encountered in
standard quantum electrodynamics.  For this reason, we had to
investigate and establish explicit representations for the
polarization vectors which play important roles in the analysis.

We have shown how to construct the Fock space of states in this
formalism. We have some interesting conclusions about the vacuum
state in this formalism, especially the energy and momentum of the
vacuum state.

Since the commutation relations for the field components are quite
complicated, we are unable to obtain general results which are
exact. We study only the electric field commutation relations. We
are able to show the effect of q-deformation for the electric
field commutation relations for different space and time points,
as a simple factor containing a modified Planck constant, only
under simplifying assumptions about polarization directions and
the assumed lack of dependence of the number operator on photon
momentum. It would certainly be worthwhile to explore this
formalism further and obtain more general results.

\end{document}